Title: Comparison of coupled nonlinear oscillator models for the transient response of power generating stations connected to low inertia systems

Authors: Marios Zarifakis, Declan Byrne, William T. Coffey, Yuri P. Kalmykov, Serguey V. Titov, Stephen J. Carrig

Published in: IEEE Transactions on Power Systems (Early Access)

URL: https://ieeexplore.ieee.org/abstract/document/8782568

Date of Publication: 31 July 2019

Print ISSN: 0885-8950

Electronic ISSN: 1558-0679

DOI: 10.1109/TPWRS.2019.2932376

Publisher: IEEE

Funding Agency: Electricity Supply Board; Science Foundation Ireland

# Comparison of coupled nonlinear oscillator models for the transient response of power generating stations connected to low inertia systems

Marios Zarifakis, Declan J. Byrne, William T. Coffey, Yuri P. Kalmykov, Serguey V. Titov and Stephen J. Carrig

*Abstract*— Coupled nonlinear oscillators, e.g., Kuramoto models, are commonly used to analyze electrical power systems. The cage model from statistical mechanics has also been used to describe the dynamics of synchronously connected generation stations. Whereas the Kuramoto model is good for describing high inertia grid systems, the cage one allows both high and low inertia grids to be modelled. This is illustrated by comparing both the synchronization time and relaxation towards synchronization of each model by treating their equations of motion in a common framework rooted in the dynamics of many coupled phase oscillators. A solution of these equations via matrix continued fractions is implemented rendering the characteristic relaxation times of a grid-generator system over a wide range of inertia and damping. Following an abrupt change in the dynamical system, the power output and both generator and grid frequencies all exhibit damped oscillations now depending on the (finite) grid inertia. In practical applications, it appears that for a small inertia system the cage model is preferable.

## I. Introduction

ELECTRIC power grids have long served as one of the principal motivations for the study of synchronization phenomena [1]. Nevertheless, the largely unexplored problem of how power generating stations *themselves* influence the dynamics of low-inertia power grids remains largely unsolved. Since that topic has become progressively more important due to increasing use of *Renewable Energy Sources* (RES), theoretical studies rooted in dynamical models of this power grid-specific problem [1],[2] are of significant interest to the energy generation industry.

Recent publications exist exploring the effect of increasing grid RES levels on its rotational inertia and dynamics following a disturbance. This has usually been achieved by finding relevant physical characteristics using simulations on various classical multimachine model testbeds, e.g., simulating the dynamic response [3]-[6], analysing the eigenvalue sensitivity [4], investigating the effects on the rate of change of rotor speed [7], and investigating inter-area power-flow oscillations (using a five-machine reduced model to represent The Western Electric Coordinating council transmission grid) [5]. Another method is Koopman mode decomposition [8],[9] which is relevant to the current paper as the nonlinear dynamic response of the system is represented as a sum of eigenfunctions in both cases, although the methods for obtaining them differ significantly.

Analysis of these systems is usually performed via a numerical solution of a many body system [3]-[7]. To obtain analytic results, however, the many body system must be reduced to a two body one, representing two coupled nonlinear phase oscillators. Here one (tagged) oscillator models the *generator* dynamics, whereas the *remaining* generators on the grid are regarded as simply a *large inertia oscillator*. The latter assumption is merely an idealized representation used to yield two tractable coupled equations. The advantage of this *Ansatz* is that the two coupled nonlinear differential equations describing the combined dynamics of the generator and grid can be solved analytically. Therefore even a relatively simple physical model may still suggest qualitative design rules for the network. This can then be used to reduce the effects of transient events. Furthermore many methods for analyzing the dynamics of a system of two coupled nonlinear oscillators exist.

Here we shall compare two physically acceptable two-body coupled oscillator models, namely, a Kuramoto-like [10]-[12] and a cage model [13],[14]. Both yield insights into the general phase behavior of a generator connected to a power grid with *finite* inertia such as prevails in Ireland.

Our solution is based on the transformation of nonlinear differential equations to an *infinite* system of differential-recurrence equations (see (A3) and (A4)). This ultimately yields the time dependent dynamic response of various characteristics of the generator and grid as sums of eigenfunctions (see (29)). Additionally, the spectra of these characteristics are described using matrix continued fractions.

One of the advantages of our method is that the results derived are analytic, yielding an intuitive understanding of the

Manuscript received xxx, 2019; revised xxx, 2019; accepted xxx, 2019. Date of publication xxx, 2019. (Corresponding author: Declan J. Byrne.)

S. V. Titov acknowledges the Electricity Supply Board for financial support.

D. J. Byrne acknowledges Science Foundation Ireland for financial support. This publication has emanated from research conducted with the financial support of Science Foundation Ireland under Grant number 17/IFB/5420.

M. Zarifakis and S. J. Carrig are with the Electricity Supply Board, Engineering and Major Projects, Dublin 3, Ireland (e-mail: marios.zarifakis@esb.ie, stephen.carrig@esb.ie).

W. T. Coffey and D. J. Byrne are with the Department of Electronic and Electrical Engineering, Trinity College, Dublin 2, Ireland (e-mail: wcoffey@mee.tcd.ie; byned12@tcd.ie).

Yu. P. Kalmykov is with the Laboratoire de Mathématiques et Physique (EA 4217), Université de Perpignan Via Domitia, F-66860, Perpignan, France (e-mail: kalmykov@univ-perp.fr).

S. V. Titov is with the Kotel'nikov Institute of Radio Engineering and Electronics of the Russian Academy of Sciences, Vvedenskii Square 1, Fryazino, Moscow Region, 141190, Russia (email: pashkin1212@yandex.ru).

effects of the nonlinear dynamics on the system. Our solution is based on the equations of motion for a low inertia grid. Hence, we are not constrained by system parameters. Moreover, as the solution completely describes the nonlinear dynamics, we can consider any size of fault. Furthermore, our method does not require a simulation to be run. Thus we believe that this method will be useful to practical engineers in the area of energy generation seeking to analyse the dynamics of low inertia grids.

## II. Equations of Motion for the Kuramoto-like Model

We first model the generator-grid interactions using Kuramoto oscillators [10]-[12]. These characterize the collective dynamics of many phase oscillator systems and have been used to analyze many dynamical systems in physics, chemistry, biology, and engineering representing coupled interacting bodies rotating or torsionally oscillating in time (e.g., chemical reactions [15], neural networks [16],[17], coupled Josephson junctions [18], optomechanical systems [19], laser arrays [20], and mean-field quantum systems [21]).

In a power grid each machine typically operates with a frequency near the grid reference frequency $\Omega = 2\pi f$ ($f = 50$ or 60 Hz). Here we describe the dynamics of the angles $\theta_j$ of the $j$th element of the power plants (generator and motor) each in accordance with a Kuramoto-like model for the phase evolution and governed by the same swing equation [15]

$$J_j \ddot{\theta}_j + K_j^K \left(\dot{\theta}_j - \Omega\right) + \sum_{l=1}^{N} \tau_{jl}^{\max} \sin\left(\theta_j - \theta_l\right) = \tau_j, \quad (1)$$

($j = 1, 2, ..., N$) for a grid consisting of $N$ units. Here $J_j$ are the moments of inertia, $K_j^K$ are the damping coefficients characterizing the damping torques of synchronous machines, $\tau_{jl}^{\max} = \tau_{lj}^{\max}$ denotes the electromagnetic torque (i.e., the coupling (interaction) strength of two machines $j$ and $l$), and $\tau_j$ is the resulting torque applied to the generator. Clearly from (1) under normal (unperturbed) conditions, all the generators in the grid operate with the same frequency $\Omega$.

We may now reduce (1) to a two body problem as discussed in the Introduction. Here a particular element, say $j = N$, is designated as a *generator*, i.e., $J_{j=N} = J_{gen}$, $\theta_{j=N} = \theta_{gen}$, $\tau_{j=N} = \tau_{gen}$, and the remaining machines are then regarded as identical, i.e., $\theta_{j\neq N} = \theta_{grid}$ and $\tau_{j\,gen}^{\max} = \tau_{gen\,j}^{\max}$. Thus we have

$$J_{gen} \ddot{\theta}_{gen} + K_{gen}^K \left(\dot{\theta}_{gen} - \Omega\right) + \sum_{l=1}^{N-1} \tau_{gen\,l}^{\max} \sin\left(\theta_{gen} - \theta_{grid}\right) = \tau_{gen}, \; j = N, \quad (2)$$

$$J_j \ddot{\theta}_{grid} + K_j^K \left(\dot{\theta}_{grid} - \Omega\right) + \tau_{j\,gen}^{\max} \sin\left(\theta_{grid} - \theta_{gen}\right) = \tau_j, \; j \neq N. \quad (3)$$

By summing all equations (3), we then have

$$\sum_{j=1}^{N-1} \left[ J_j \ddot{\theta}_{grid} + K_j^K \left(\dot{\theta}_{grid} - \Omega\right) + \tau_{j\,gen}^{\max} \sin\left(\theta_{grid} - \theta_{gen}\right) \right] = \sum_{j=1}^{N-1} \tau_j. \quad (4)$$

Equations (2) and (4) can now be rewritten in the simultaneous form:

$$J_{gen} \ddot{\theta}_{gen} + K_{gen}^K \left(\dot{\theta}_{gen} - \Omega\right) + \tau_{el} \sin\left(\theta_{gen} - \theta_{grid}\right) = \tau_{gen}, \quad (5)$$

$$J_{grid} \ddot{\theta}_{grid} + K_{grid}^K \left(\dot{\theta}_{grid} - \Omega\right) + \tau_{el} \sin\left(\theta_{grid} - \theta_{gen}\right) = \tau_{grid}, \quad (6)$$

where

$$J_{grid} = \sum_{j=1}^{N-1} J_j, \; K_{grid}^K = \sum_{j=1}^{N-1} K_j^K, \; \tau_{grid} = \sum_{j=1}^{N-1} \tau_j,$$

$$\tau_{el} = \sum_{j=1}^{N-1} \tau_{j\,gen} = \sum_{j=1}^{N-1} \tau_{gen\,j}.$$

Next, we rewrite (5) and (6) in the generic form

$$\ddot{\theta}_i + \beta_i^K \left(\dot{\theta}_i - \Omega\right) + \xi_i \sin\left(\theta_i - \theta_j\right) = \bar{\tau}_i, \quad (7)$$

where, $i = grid, gen$, $j = gen, grid$, $\xi_i = \tau_{el}/J_i$, $\bar{\tau}_i = \tau_i/J_i$, and in this case, $\beta_i^K = K_i^K/J_i$. If the condition

$$J_{grid}/J_{gen} = K_{grid}^K/K_{gen}^K = N \quad (8)$$

is imposed (i.e., the grid consists of $N$ identical generators) then by subtraction of (7) when $i=gen$ from (7) when $i=grid$ we have the evolution equation for the rotor angle (namely the equation of motion of a *driven damped pendulum*), viz.,

$$\ddot{\delta} + \beta \dot{\delta} + \xi \sin \delta = \tau, \quad \delta = \theta_{grid} - \theta_{gen}, \quad (9)$$

where $\xi = \xi_{grid} + \xi_{gen}$, $\tau = \bar{\tau}_{grid} - \bar{\tau}_{gen}$, and in this case $\beta = \beta_{grid}^K = \beta_{gen}^K = K_{gen}^K/J_{gen} = K_{grid}^K/J_{grid}$. This pendulum model is also commonly used for the dynamic response of a synchronous generator in an *infinite* grid [22],[23]. Now, to describe the effects of *finite* grid inertia we introduce a new variable $x$, namely the *ratio of the grid inertia to the generator inertia*, $x = J_{grid}/J_{gen}$ so that in (9)

$$\xi = \tau_{el} \left(1 + x^{-1}\right)/J_{gen}, \quad \tau = -\tau_{gen} \left(1 + x^{-1}\right)/J_{gen}, \quad (10)$$

and $x$ has no effect on $\beta$ for the Kuramoto-like model. Equation (9) can be solved using the matrix continued fraction solution of the damped pendulum equation (see Appendix), yielding the general solution of (7) for arbitrary $\beta_i^K$, $\xi_i$ and $\bar{\tau}_i$.

We now seek the evolution equation for the grid and generator angles in terms of the rotor angle. Here one must in general solve the (simultaneous) system of (5) and (6). However, by imposing the condition proposed in (8), the solution simplifies. First by adding (5) and (6), we have

$$J_{grid} \ddot{\theta}_{grid} + J_{gen} \ddot{\theta}_{gen} + K_{grid}^K \dot{\theta}_{grid} + K_{gen}^K \dot{\theta}_{gen} = \left(K_{grid}^K + K_{gen}^K\right)\Omega \quad (11)$$

Solving (11) subject to the condition (8) and with $\dot{\theta}_i(0) = \Omega$, we have

$$J_{grid} \ddot{\theta}_{grid} + J_{gen} \ddot{\theta}_{gen} = 0. \quad (12)$$

Then by integrating (12) twice subject to the initial condition $\dot{\theta}_i(0) = \Omega$, corresponding to *unperturbed* (i.e., steady) rotation of grid and generator, we have

$$J_{grid} \left(\theta_{grid}(t) - \theta_{grid}(0)\right) + J_{gen} \left(\theta_{gen}(t) - \theta_{gen}(0)\right) = (J_{grid} + J_{gen})\Omega t, \quad (13)$$

so that the time evolution of the angles $\theta_i(t)$ is explicitly

$$\theta_{grid}(t) = \Omega t + \frac{J_{grid}\theta_{grid}(0) + J_{gen}\theta_{gen}(0)}{J_{grid} + J_{gen}} + \frac{J_{gen}}{J_{grid} + J_{gen}}\delta(t), \quad (14)$$

$$\theta_{gen}(t) = \Omega t + \frac{J_{grid}\theta_{grid}(0) + J_{gen}\theta_{gen}(0)}{J_{grid} + J_{gen}} - \frac{J_{grid}}{J_{grid} + J_{gen}}\delta(t). \quad (15)$$

## III. EQUATIONS OF MOTION FOR THE CAGE MODEL

For comparison, we also analyze the grid and generator dynamics using the alternative cage model [13],[14], where the dynamics of the angles $\theta_j$ are described by the many body equation

$$J_j \ddot{\theta}_j + \sum_{l=1}^{N} K_{jl}^C \left( \dot{\theta}_j - \dot{\theta}_l \right) + \sum_{l=1}^{N} \tau_{jl}^{\max} \sin\left( \theta_j - \theta_l \right) = \tau_j. \quad (16)$$

The essential difference between (16) and (1) pertaining to the Kuramoto-like model is that there [12] the damping torque is proportional to the *deviation* of the grid/generator frequency from the reference frequency $\Omega$, namely $K_j^K \left( \dot{\theta}_j - \Omega \right)$, whereas in the cage model [2],[13] the damping torque of the *j*-th generator (say) is proportional to the *difference* between its particular frequency and all other generator frequencies, namely $\sum_{l=1}^{N} K_{jl}^C \left( \dot{\theta}_j - \dot{\theta}_l \right)$. The cage (itinerant oscillator) model is widely used in chemical physics (for a detailed summary of its applications see [14] and references therein).

Following the method used to analyze the Kuramoto-like case, we again reduce (16) to a two body system so that

$$J_{gen} \ddot{\theta}_{gen} + \sum_{l=1}^{N-1} K_{genl}^C (\dot{\theta}_{gen} - \dot{\theta}_{grid}) + \sum_{l=1}^{N-1} \tau_{genl}^{\max} \sin(\theta_{gen} - \theta_{grid}) = \tau_{gen}, \quad (17)$$

$$\sum_{j=1}^{N-1} \left[ J_j \ddot{\theta}_{grid} + K_{jgen}^C \left( \dot{\theta}_{grid} - \dot{\theta}_{gen} \right) + \tau_{jgen}^{\max} \sin\left( \theta_{grid} - \theta_{gen} \right) \right] = \sum_{j=1}^{N-1} \tau_j. \quad (18)$$

Then via the substitutions $K^C = \sum_{l=1}^{N-1} K_{genl}^C = \sum_{j=1}^{N-1} K_{jgen}^C$, $J_{grid} = \sum_{j=1}^{N-1} J_j$, $\tau_{grid} = \sum_{j=1}^{N-1} \tau_j$, $\tau_{el} = \sum_{l=1}^{N-1} \tau_{genl}^{\max} = \sum_{j=1}^{N-1} \tau_{jgen}^{\max}$, we have the equations of motion rendered as

$$J_{gen} \ddot{\theta}_{gen} + K^C \left( \dot{\theta}_{gen} - \dot{\theta}_{grid} \right) + \tau_{el} \sin\left( \theta_{gen} - \theta_{grid} \right) = \tau_{gen}, \quad (19)$$

$$J_{grid} \ddot{\theta}_{grid} + K^C \left( \dot{\theta}_{grid} - \dot{\theta}_{gen} \right) + \tau_{el} \sin\left( \theta_{grid} - \theta_{gen} \right) = \tau_{grid}. \quad (20)$$

We now rewrite (19) and (20) as

$$\ddot{\theta}_i + \beta_i^C \left( \dot{\theta}_i - \dot{\theta}_j \right) + \xi_i \sin\left( \theta_i - \theta_j \right) = \overline{\tau}_i, \quad (21)$$

where $\xi_i$ and $\overline{\tau}_i$ are defined as in (7) but in this case $\beta_i^C = K^C / J_i$. Then by subtraction of (21) when *i*=*gen* from (21) when *i*=*grid* we now have (9) where in this instance $\beta = \beta_{grid}^C + \beta_{gen}^C$, or in terms of the grid to generator inertia ratio,

$$\beta = K^C \left( 1 + x^{-1} \right) / J_{gen}. \quad (22)$$

Equation (22) is very important because it emphasizes the essential difference between the cage and Kuramoto-like models for finite inertia grids.

To find the evolution equation for the grid and generator angles in terms of the rotor angle, we follow the method for the Kuramoto-like model. By addition of (19) and (20) we again have (12), thus leading to (14) and (15).

## IV. TRANSIENT RESPONSE FUNCTION

Supposing that a disturbance to the motion occurs at the instant $t = 0$ whereupon the interaction parameter $\tau_{el}$ alters from an initial value $\tau_{el}^I$ to $\tau_{el}^{II}$ so that we require the transient behavior, starting from an equilibrium state I with rotor angle $\delta(0) = \delta_I$ to a new equilibrium state II with $\delta(t \to \infty) = \delta_{II}$. Both $\delta_I$ and $\delta_{II}$ depend on the applied torque of the turbine because $\sin \delta_I = \tau_{gen} / \tau_{el}^I = -\tau_{grid} / \tau_{el}^I$ and $\sin \delta_{II} = \tau_{gen} / \tau_{el}^{II} = -\tau_{grid} / \tau_{el}^{II}$. The initial value of $\dot{\delta}_I = \dot{\delta}(0)$ is zero. To find $\delta(t)$, we introduce the complete set of functions

$$a_{nq}(t) = \dot{\delta}^n \sin^q \left( \delta - \delta_{II} \right), \quad (23)$$

$$b_{nq}(t) = \dot{\delta}^n \sin^q \left( \delta - \delta_{II} \right) \left[ 1 - \cos\left( \delta - \delta_{II} \right) \right]. \quad (24)$$

Thus, $\delta(t)$ can then be expressed in terms of $a_{nq}(t)$ and $b_{nq}(t)$ as

$$\cos \delta(t) = (1 - b_{00}(t)) \cos \delta_{II} - a_{01}(t) \sin \delta_{II}, \quad (25)$$

or

$$\delta(t) = \int a_{01}(t') dt'. \quad (26)$$

We have (v. Appendix) from (9) a set of differential-recurrence equations governing the evolution of $a_{nq}(t)$ and $b_{nq}(t)$ (see (A3) and (A4)). The behavior of any selected function is coupled to that of all the others, so generating an *infinite hierarchy* of equations.

Numerical solutions for $a_{nq}(t)$ and $b_{nq}(t)$ may be obtained via matrix continued fractions (v. Appendix) or else via direct diagonalization by writing (A3) and (A4) as a first-order matrix differential equation, viz.,

$$\frac{d}{dt} \mathbf{C} + \mathbf{X} \mathbf{C} = 0. \quad (27)$$

Here $\mathbf{C}$ is the super column vector with elements comprised of the vectors $\mathbf{C}_n$ and $\mathbf{X}$ is the time independent infinite system matrix defined as

$$\mathbf{C} = \begin{pmatrix} \mathbf{C}_1 \\ \mathbf{C}_2 \\ \vdots \end{pmatrix}, \quad \mathbf{X} = -\begin{pmatrix} \mathbf{Q}_1 & \mathbf{Q}_1^+ & 0 & 0 & \cdots \\ \mathbf{Q}_2^- & \mathbf{Q}_2 & \mathbf{Q}_2^+ & 0 & \cdots \\ 0 & \mathbf{Q}_3^- & \mathbf{Q}_3 & \mathbf{Q}_3^+ & \cdots \\ 0 & 0 & \mathbf{Q}_4^- & \mathbf{Q}_4 & \cdots \\ \vdots & \vdots & \vdots & \vdots & \ddots \end{pmatrix},$$

where the matrices $\mathbf{Q}_n$ and $\mathbf{Q}_n^\pm$ and the vector $\mathbf{C}_n$ are given by (A6) in the Appendix. Equation (27) can then be solved using the standard inversion methods of linear algebra, i.e., by progressively increasing the size $L$ of $\mathbf{X}$ until convergence is attained. We then have

$$\mathbf{C}(t) = e^{-\mathbf{X}t} \mathbf{C}(0) = \mathbf{U} e^{-\mathbf{\Lambda}t} \mathbf{U}^{-1} \mathbf{C}(0), \quad (28)$$

where $\mathbf{\Lambda} = \mathbf{U}^{-1} \mathbf{X} \mathbf{U}$ is a diagonal matrix with elements composed of all the eigenvalues $\lambda_k$ of $\mathbf{X}$ and $\mathbf{U}$ reduces $\mathbf{X}$ to diagonal form. The columns of $\mathbf{U}$ are the components of the column eigenvectors of the matrix $\mathbf{X}$. Equation (28) can then be used to find $\delta(t)$ and $\dot{\delta}(t)$, viz.,

$$\delta(t) = \sum_j^L c_j e^{-\lambda_j t} + \delta_{II}, \quad \dot{\delta}(t) = \sum_j^L d_j e^{-\lambda_j t}, \quad (29)$$

where $c_j$ and $d_j$ are the elements of the vectors (v. (26) and (28))

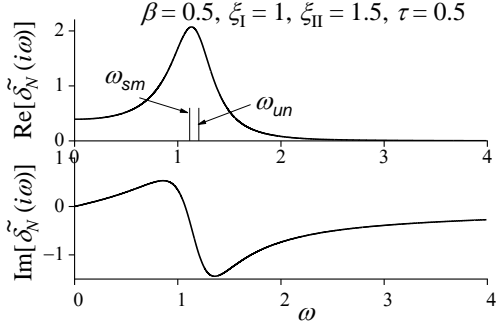

Fig. 1. Frequency dependence of the real and imaginary parts of $\tilde{\delta}_N(i\omega)$, $\delta_N(t) = (\delta(t) - \delta_{II})/(\delta_I - \delta_{II})$, (32), for damping parameter $\beta = 0.5$, torque parameter $\tau = 0.5$, initial coupling parameter $\xi_I = 1$, and final coupling parameter $\xi_{II} = 1.5$.

$$\mathbf{c} = (\mathbf{U}_{2q_{max}+3}) \circ (\mathbf{\Lambda}^{-1}\mathbf{U}^{-1}\mathbf{C}(0))^T, \quad \mathbf{d} = (\mathbf{U}_{2q_{max}+3}) \circ (\mathbf{U}^{-1}\mathbf{C}(0))^T, \quad (30)$$

with $\mathbf{U}_{2q_{max}+3}$ denoting row $2q_{max}+3$ of $\mathbf{U}$ ($q_{max}$ is defined in the Appendix).

To analyze the dynamics of system immediately following an alteration of $\tau_{el}$ via Fourier transformation, we now introduce the *normalized* rotor angle,

$$\delta_N(t) = \frac{\delta(t) - \delta_{II}}{\delta_I - \delta_{II}}, \quad (31)$$

describing the evolution of $\delta(t)$ from a state I, where $\delta_N(0) = 1$, to the stationary state II, where $\delta_N(t \to \infty) = 0$, after the removal of the disturbance. Analysis of $\delta_N(t)$ is preferable in comparison to that of $\delta(t)$ as it is already normalized, viz. $\delta_N(0) = 1$, and has finite area under the decay curve. Thus its spectrum (see Fig.1) has no *singular* points so that Fourier transformation may be used to analyze it. By one-sided Fourier transformation the set of differential-recurrence evolution equations for $a_{nq}(t)$ and $b_{nq}(t)$ is then converted to a set of algebraic equations for the one-sided Fourier transforms $\tilde{a}_{nq}(i\omega)$ and $\tilde{b}_{nq}(i\omega)$ (see Appendix). These are solved via matrix continued fractions so yielding a formal analytical result for the one-sided Fourier transform of $\delta_N(t)$, viz.,

$$\tilde{\delta}_N(i\omega) = \lim_{s \to i\omega} \int_0^\infty \delta_N(t) e^{-st} dt = \frac{1}{i\omega}\left(\frac{\tilde{a}_{10}(i\omega)}{\delta_I - \delta_{II}} + 1\right). \quad (32)$$

$\delta_N(t)$ can then be recovered from $\tilde{\delta}_N(i\omega)$ via inverse Fourier transformation, v. [2].

Figs. 1 and 2 show the spectrum $\tilde{\delta}_N(i\omega)$ and $\delta(t)$.

V. CHARACTERISTIC FREQUENCIES AND TIMES

If $\delta(t)$ represents a decaying oscillatory function as exhibited in Fig. 2, the frequency of the oscillations $\omega_\delta$ of $\delta(t)$ can be roughly estimated either from the (linear) response for small disturbances [24] as

$$\omega_{sm} = \sqrt{\xi\cos\delta_{II} - \beta^2/4}, \quad (33)$$

or from the undamped pendulum equation $\ddot{\delta} + \xi\sin\delta = 0$

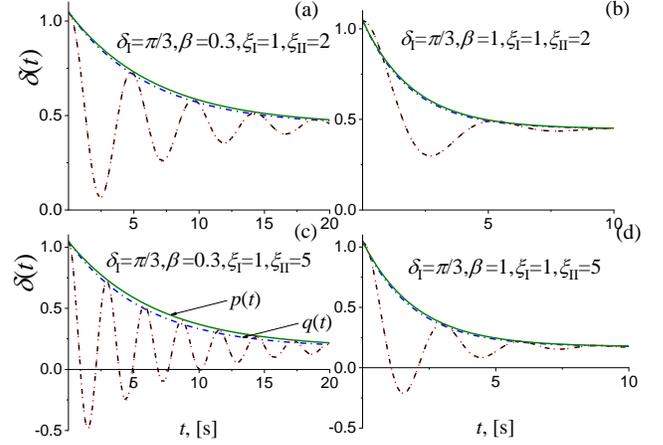

Fig. 2. (Color online) Time dependence of the function $\delta(t)$ for initial coupling parameter $\xi_I = 1$ and initial angle $\delta_I = \pi/3$ and various values of damping parameters $\beta$ and final coupling parameters $\xi_{II}$. Dashed lines: $\delta(t)$ via (29). Dotted lines: numerical solution of (9). Dashed-dotted lines: exponential approximation for the main maximum of $\delta(t)$, viz., $q(t) = (\delta_I - \delta_{II})e^{-t/T_{int}} + \delta_{II}$ with $T_{int}$ given by (36). Solid lines: linear response $p(t) = (\delta_I - \delta_{II})e^{-\beta t/2} + \delta_{II}$.

($\beta = \tau = 0$) yielding [2]

$$\omega_{un} = \frac{\pi\sqrt{\xi}}{2K(m)}, \quad (34)$$

where $K(m)$ is a complete elliptic integral of the first kind [25] with modulus $m = \sin^2(\delta_0/2)$ and $\delta_0$ is the amplitude. The frequency may also be extracted from the spectrum of $\delta(t)$ (see Fig. 1). Fig. 3 shows the behavior of $\omega_\delta$ vs. the parameter $\xi_{II}$ corresponding to the final state II. Clearly both distinct methods of estimating the frequency yield similar results.

Now, *synchronization dynamics* in power-grid networks is of great practical significance. For example, synchronization of all power generators in the same interconnection is an absolute requirement for a power grid to operate. One of the parameters characterizing the synchronization process is the *synchronization time*, i.e., the time needed to *recover synchronized operation* of the grid following a *desynchronization event*. Thus, we shall use the concept of the integral relaxation time of a relaxation function (from statistical mechanics) to estimate the time scale of a synchronization process. We begin with the *Ansatz* that a given normalized function $q_N(t) = e^{-t/T}$ has exponential behavior with relaxation time $T$, so that the integral relaxation time may be rigorously defined as

$$T_{int} = \int_0^\infty q_N(t)dt = T. \quad (35)$$

However, we may also introduce $T_{int}$ for an *arbitrary normalized response function* $\delta_N(t)$ which in general *is not exponential* [13]. This time characterizes the general process whereby a physical system, following changes in a particular system parameter, evolves from one stationary state to another stationary one and therefore may be regarded as a *characteristic time* of a *synchronization process*. Thus if $\delta_N(t)$ simultaneously oscillates and decays, we can approximate the curve by the envelope crossing all *positive* maxima as the exponential $q_N(t) = e^{-t/T}$. Then the integral relaxation time can be approximately evaluated via the *time to reach the first maximum* $\delta_N(T_{os}) = q_N(T_{os}) = e^{-T_{os}/T_{int}}$ at $t = T_{os}$ so that

$$T_{int} = T = -T_{os} / \ln \delta_N(T_{os}). \qquad (36)$$

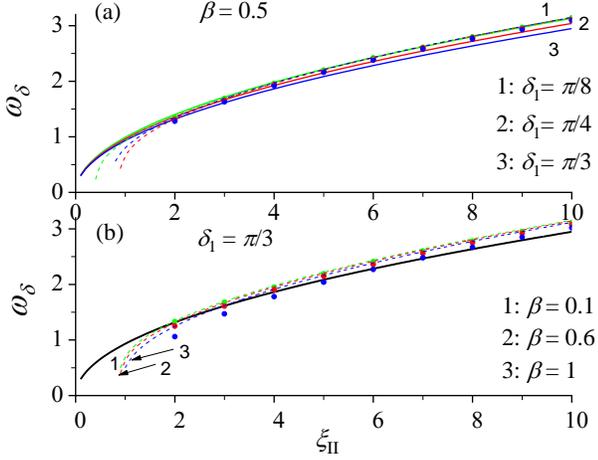

Fig. 3. (Color online) Frequency of oscillation of the relaxation function $\delta(t)$ vs. the final coupling parameter $\xi_{II}$ for initial coupling parameter $\xi_I = 1$ and various initial angles (a) $\delta_I = \pi/8, \pi/4, \pi/3$ and damping parameters (b) $\beta = 0.1, 0.6, 1$. Solid line: $\omega_{un}$, (34), dashed line: $\omega_{sm}$, (33) and circles: frequency of maximum value of $\mathrm{Re}[\tilde{\delta}_N(i\omega)]$ (see Fig. 1).

For a *linear* response the time of the first maximum is simply the time interval between two adjacent maxima, is given by

$$T_{os} = 2\pi / \omega_\delta, \qquad (37)$$

where $\omega_\delta$ may be approximated via (33), (34) or else from the spectrum of $\delta_N(t)$ (see Figs. 1 and 3). Now, for a nonlinear response $T_{os}$ can be evaluated exactly using (29) from the smallest nonzero root of $\dot{\delta}(t) = 0$, i.e.,

$$\sum_j^L d_j e^{-\lambda_j t} = 0. \qquad (38)$$

The curve crossing all positive maxima of $\delta(t)$ is then $q(t) = (\delta_I - \delta_{II})e^{-t/T_{int}} + \delta_{II}$. For the linear response this curve can be explicitly expressed as [24] $p(t) = (\delta_I - \delta_{II})e^{-\beta t/2} + \delta_{II}$.

## VI. RESULTS AND DISCUSSION

Fig. 2 shows the time dependence of $\delta(t)$ via (29) for initial coupling $\xi_I = 1$ and initial angle $\delta_I = \pi/3$ and various values of damping $\beta$ and final coupling $\xi_{II}$. By using (36) we have a rather accurate exponential approximation, viz., $q(t) = (\delta_I - \delta_{II})e^{-t/T_{int}} + \delta_{II}$ for the main maximum of $\delta(t)$. Thus $T_{int}$ can be safely regarded in this instance as a characteristic time of the relaxation process. Fig. 4 shows the integral relaxation time extracted via the exponential approximation for the main maximum of the relaxation function. Clearly for a *highly nonlinear* situation (large $|\xi_{II} - \xi_I|$ and small $\beta$) the approximation (37) fails to describe the relaxation effectively.

To model the effects of *finite* grid inertia we use the definitions of the system parameters in terms of the grid to generator inertia ratio $x$, (10) and for the cage model (22). The ratio $x$ has no effect on $\beta$ for the Kuramoto-like model. Figs. 5-7 show the time dependence of $\delta(t)$, $\omega_{gen}(t)$, and $\omega_{grid}(t)$ for various $x$ using both the cage and Kuramoto-like models. Here $\delta(t)$ is calculated via (29) and $\omega_{gen} = \dot{\theta}_{gen}, \omega_{grid} = \dot{\theta}_{grid}$ are calculated via (14), (15), and (29). The main effect of the low inertia grid is an increase in the frequency of the oscillations (which can be intuitively predicted from (10), (33), and (34)). Additionally, the amplitudes of the oscillations of $\delta(t)$ and $\omega_{gen}(t)$ for the cage model are reduced for low inertia (as expected from (22)). For a small, virtually isolated grid like that of Ireland where $x \sim 10$, these results are highly relevant.

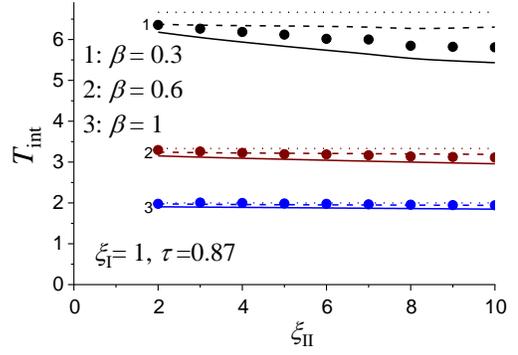

Fig. 4. (Color online) Integral relaxation time $T_{int}$, (36), of the function $q(t)$ vs. final coupling parameter $\xi_{II}$ for torque parameter $\tau = 0.87$ initial coupling parameter $\xi_I = 1$ and various damping parameters $\beta = 0.3, 0.6, 1$. Solid lines: $\omega_\delta = \omega_{sm}$, (33); circles: $\omega_\delta = \omega_{max}$, where $\omega_{max}$ is frequency of maximum value of $\mathrm{Re}[\tilde{\delta}_N(i\omega)]$ (see Fig. 1); dashed lines: $T_{os}$ via (38); dotted lines: linear response, $T_{int} = 2/\beta$ [24].

## VII. CONCLUSIONS

We have presented a comparative study of the electromechanical dynamics of a generator connected to a low-inertia power grid using coupled phase oscillators in the form of the cage and Kuramoto models. We conclude that both yield comparable results, which may serve as analytical guidelines for qualitative studies of the dynamical response of a generator connected for a low inertia transmission system.

This paper also extends our previous work on energy grid dynamics [2] incorporating many improvements. For example, [2] only considers the dynamics described by the cage model, whereas here we also treat the dynamics via the Kuramoto-like model (commonly used to analyze grid systems [10]-[12]). Also we have given a new description of the time dependent dynamics derived in an analytic manner directly from the system matrix describing the generator and grid (see (27)-(30)). Moreover the effects of low grid inertia are now shown explicitly in the equations via a *grid to generator inertia ratio* (see (10) and (22)). We have also provided analytic equations describing the dynamics of the grid and generator frequencies (see (14), (15), and (29)). Furthermore we have introduced different methods of describing both the characteristic frequencies and times of the grid-generator dynamics following a disturbance (see Section 5).

The advantage of the Kuramoto-like model is that the damping on a generator can be calculated directly from the generator rotor frequency and so does not require knowledge of the dynamics of the grid. This is appropriate if all the generators in the grid have *identical* frequencies (i.e., frequency is synchronized) and this frequency is constant (and chosen as the reference frequency). This constraint is almost fulfilled in a large-scale power grid (e.g., the European grid) with high inertia, where the *common frequency is so tightly controlled as to be very near the nominal frequency*. Then using the Kuramoto-like model with the nominal frequency as the reference frequency is justified. However, the disadvantage of using this model is that it is inappropriate for analysing *small inertia systems* where the grid dynamics cannot be so tightly controlled. The advantage of the cage model in this instance is that it does not rely at all on the assumption that the grid has a constant reference frequency and therefore is appropriate for analysing these small inertia systems. Moreover, it can also be used to analyze *high inertia* grid systems because for infinite grid inertia the damping to inertia ratio coefficient $\beta$ defined for the cage model, $\beta = K^C(J_{grid}^{-1} + J_{gen}^{-1})$, simply reduces to $\beta = K^C/J_{gen}$ which is equivalent to the Kuramoto-like definition. In a typical energy grid $J_{grid} \gg J_{gen}$ so that both solutions will not significantly differ.

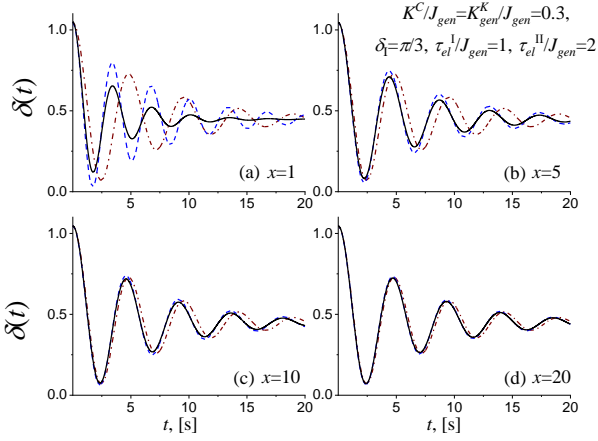

Fig. 5. (Color online) Time dependence of the angle $\delta(t)$ for the cage (solid lines) and Kuramoto-like (dashed lines) models and an infinite grid inertia system (dashed-dotted line) for $K^C/J_{gen} = K^K_{gen}/J_{gen} = 0.3$, $\delta_\mathrm{I} = \pi/3$, $\tau^\mathrm{I}_{el}/J_{gen} = 1$, $\tau^\mathrm{II}_{el}/J_{gen} = 2$ and various values of the grid inertia to generator

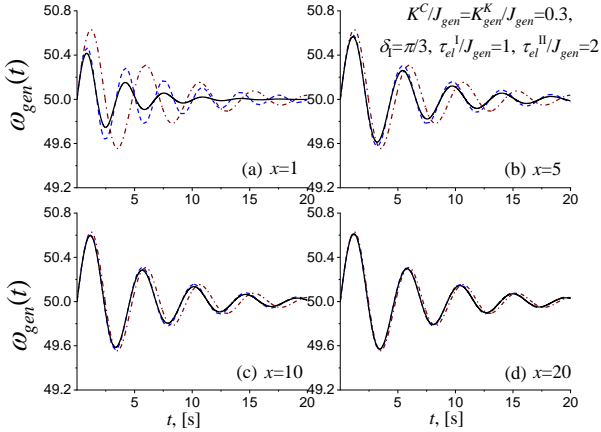

Fig. 6. (Color online) Time dependence of the generator frequency $\omega_{gen}(t)$ for the cage (solid line) and Kuramoto-like (dashed line) models and an infinite grid inertia system (dashed-dotted line) for $K^C/J_{gen} = K^K_{gen}/J_{gen} = 0.3$, $\delta_\mathrm{I} = \pi/3$, $\tau^\mathrm{I}_{el}/J_{gen} = 1$, $\tau^\mathrm{II}_{el}/J_{gen} = 2$ and various values of the grid inertia to generator inertia ratio $x$.

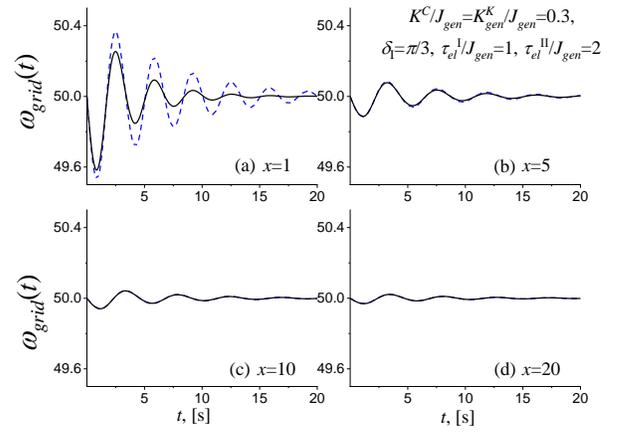

Fig. 7. (Color online) Time dependence of the grid frequency $\omega_{grid}(t)$ for the cage (solid lines) and Kuramoto-like (dashed lines) models for $K^C/J_{gen} = K^K_{gen}/J_{gen} = 0.3$, $\delta_\mathrm{I} = \pi/3$, $\tau^\mathrm{I}_{el}/J_{gen} = 1$, $\tau^\mathrm{II}_{el}/J_{gen} = 2$ and various values of the grid inertia to generator inertia ratio $x$.

We have demonstrated how the two degree of freedom dynamical equations corresponding to both models may be solved analytically. The advantage of such solutions over numerical integration methods is that they yield results in closed form, allowing one to transparently analyze the relation between the *input* variables and the *response*. We have also proposed an accurate technique for calculating the characteristic times of both models over a wide range of inertia and friction. The analytical solution so yielded may serve as a basis for a qualitative understanding of how actual generators behave following a disturbance resulting in a grid-frequency change as well as revealing more general features of complex multi-degree of freedom systems. They could also be extended to two or more grid connected generators, or to consider a system of non-identical generators (by averaging over appropriate distribution functions), yielding quantitatively better results.

Thus we have demonstrated how lowering the inertia of a grid (e.g., by increasing the proportion of renewable energy sources) may lead to undesirable behavior (see Fig. 7). Therefore we must also consider methods to counteract such effects, e.g., using Power System Stabilizers (PSS) and Automatic Voltage Regulators (AVR) to dampen the oscillations on the grid. Recently [26] we explored this using the conventional IEEE PSS1A, where the nonlinear grid-generator dynamics derived in this paper (see (29)) are used as inputs to the PSS and AVR.

APPENDIX

*A. Matrix Continued Fraction Solution of (9)*

We rewrite (9) as
$$\ddot{\delta} = -\beta\dot{\delta} - \xi_\mathrm{II}\left\{\sin(\delta - \delta_\mathrm{II})\cos\delta_\mathrm{II} + [\cos(\delta - \delta_\mathrm{II}) - 1]\sin\delta_\mathrm{II}\right\}, \quad (A1)$$
where $\xi_\mathrm{II} = \xi = \xi_{grid} + \xi_{gen}$ and $\xi_\mathrm{II}\sin\delta_\mathrm{II} = \tau = \bar{\tau}_{grid} - \bar{\tau}_{gen}$. The time derivatives of the functions $a_{nq}(t)$ given by (23) are

$$\frac{d}{dt}a_{nq}(t) = n\dot{\delta}^{n-1}\ddot{\delta}\sin^q(\delta-\delta_{II}) + q\dot{\delta}^n\sin^{q-1}(\delta-\delta_{II})\cos(\delta-\delta_{II}).$$
(A2)

By substituting (A1) into (A2), we then have differential-recurrence relations for the $a_{nq}(t)$, viz.,

$$\frac{d}{dt}a_{nq} = -\beta n a_{nq} + q a_{n+1q-1} - q b_{n+1q-1}$$
$$-\xi_{II}n\left[a_{n-1q+1}\cos\delta_{II} - b_{n-1q}\sin\delta_{II}\right],$$
(A3)

whence

$$\frac{d}{dt}b_{nq} = -\beta n b_{nq} + (q+1)a_{n+1q+1} - q b_{n+1q-1}$$
$$-\xi_{II}n\left[a_{n-1q+2}\sin\delta_{II} + b_{n-1q+1}\cos\delta_{II} - 2b_{n-1q}\sin\delta_{II}\right].$$
(A4)

Equations (A3) and (A4) can be transformed into the tri-diagonal matrix form

$$\frac{d}{dt}\mathbf{C}_n(t) = \mathbf{Q}_n^-\mathbf{C}_{n-1}(t) + \mathbf{Q}_n\mathbf{C}_n(t) + \mathbf{Q}_n^+\mathbf{C}_{n+1}(t)$$
(A5)

with the initial conditions

$$\mathbf{C}_1(0) = \begin{pmatrix} 1 \\ 1-\cos(\delta_I - \delta_{II}) \\ \sin(\delta_I - \delta_{II}) \\ \sin(\delta_I - \delta_{II})[1-\cos(\delta_I - \delta_{II})] \\ \sin^2(\delta_I - \delta_{II}) \\ \sin^2(\delta_I - \delta_{II})[1-\cos(\delta_I - \delta_{II})] \\ \vdots \end{pmatrix},$$

$$\mathbf{C}_n(0) = 0, \; n > 2.$$

Here the infinite column vectors $\mathbf{C}_n$ and the matrices $\mathbf{Q}_n$ and $\mathbf{Q}_n^\pm$ are

$$\mathbf{C}_n = \begin{pmatrix} a_{n-10} \\ b_{n-10} \\ a_{n-11} \\ b_{n-11} \\ \vdots \end{pmatrix}, \mathbf{Q}_n^+ = \begin{pmatrix} 0 & 0 & 0 & 0 & 0 & \cdots \\ 0 & 0 & 1 & 0 & 0 & \cdots \\ 1 & -1 & 0 & 0 & 0 & \cdots \\ 0 & -1 & 0 & 0 & 2 & \cdots \\ 0 & 0 & 2 & -2 & 0 & \cdots \\ \vdots & \vdots & \vdots & \vdots & \vdots & \ddots \end{pmatrix}, \mathbf{Q}_n = \beta(1-n)\mathbf{I},$$

$$\mathbf{Q}_n^- = \xi_{II}(1-n)\begin{pmatrix} 0 & -\sin\delta_{II} & \cos\delta_{II} & 0 & 0 & \cdots \\ 0 & -2\sin\delta_{II} & 0 & \cos\delta_{II} & \sin\delta_{II} & \cdots \\ 0 & 0 & 0 & -\sin\delta_{II} & \cos\delta_{II} & \cdots \\ 0 & 0 & 0 & -2\sin\delta_{II} & 0 & \cdots \\ \vdots & \vdots & \vdots & \vdots & \vdots & \ddots \\ \vdots & \vdots & \vdots & \vdots & \vdots & \ddots \end{pmatrix}$$
(A6)

and $\mathbf{I}$ is the unit matrix of infinite dimension.

Equation (A5) may be solved in the frequency domain via matrix continued fractions [13],[27]. Using the initial conditions $\delta(0) = \delta_I$ and $\dot{\delta}(0) = 0$ and taking the Laplace transformation of (A5), we have the following matrix recurrence relations for $\tilde{\mathbf{C}}_n(s)$

$$-s\tilde{\mathbf{C}}_1(s) + \mathbf{Q}_1^+\tilde{\mathbf{C}}_2(s) = -\mathbf{C}_1(0),$$
(A7)

$$\mathbf{Q}_n^-\tilde{\mathbf{C}}_{n-1}(s) + (\mathbf{Q}_n - s\mathbf{I})\tilde{\mathbf{C}}_n(s) + \mathbf{Q}_n^+\tilde{\mathbf{C}}_{n+1}(s) = 0, \; (n > 1). \quad (A8)$$

The solution of (A7) and (A8) for $\tilde{\mathbf{C}}_n(s)$ is

$$\tilde{\mathbf{C}}_1(s) = \mathbf{\Delta}_1\mathbf{C}_1(0),$$
(A9)

where $\mathbf{\Delta}_n$ is defined by the recurrence equation

$$\mathbf{\Delta}_n = \left[s\mathbf{I} - \mathbf{Q}_n - \mathbf{Q}_n^+\mathbf{\Delta}_{n+1}\mathbf{Q}_{n+1}^-\right]^{-1},$$
(A10)

so representing $\mathbf{\Delta}_n$ itself as an infinite matrix continued fraction [13].

Our solution is very amenable to computation (various algorithms for calculating matrix continued fractions are discussed in [13]). As far as the calculation of the (infinite) $\mathbf{\Delta}_n$ is concerned, we first approximate it by some matrix continued fraction of finite order (by putting $\mathbf{Q}_{n+1} = 0$ at some $n = n_{max}$). Simultaneously, we confined the dimensions of $\mathbf{Q}_n$ and $\mathbf{Q}_n^\pm$ to some finite number $2(q_{max}+1)$. Both $n_{max}$ and $q_{max}$ depend on the model parameters selected and must also be chosen taking account of the desired degree of accuracy of the calculation.

We note that the solution of the damped pendulum equation (9) in terms of continued fractions has been given in [2] via an *alternative* complete set of relaxation functions. However, the new functions $a_{nq}(t)$ and $b_{nq}(t)$ have the significant advantages that they are both real, have finite area under their curves in the time domain and that their Fourier transforms have no singular points.